\documentstyle[12pt]{article}
\def\PRL #1 #2 #3{{\sl Phys. Rev. Lett.} {\bf#1} (#2) #3}
\def\NPB #1 #2 #3{{\sl Nucl. Phys.} {\bf B#1} (#2) #3}
\def\NPBFS #1 #2 #3 #4{{\sl Nucl. Phys.} {\bf B#2} [FS#1] (#3) #4}
\def\CMP #1 #2 #3{{\sl Commun. Math. Phys.} {\bf #1} (#2) #3}
\def\PRD #1 #2 #3{{\sl Phys. Rev.} {\bf D#1} (#2) #3}
\def\PLA #1 #2 #3{{\sl Phys. Lett.} {\bf #1A} (#2) #3}
\def\PLB #1 #2 #3{{\sl Phys. Lett.} {\bf #1B} (#2) #3}
\def\JMP #1 #2 #3{{\sl J. Math. Phys.} {\bf #1} (#2) #3}
\def\PTP #1 #2 #3{{\sl Prog. Theor. Phys.} {\bf #1} (#2) #3}
\def\SPTP #1 #2 #3{{\sl Suppl. Prog. Theor. Phys.} {\bf #1} (#2) #3}
\def\AoP #1 #2 #3{{\sl Ann. of Phys.} {\bf #1} (#2) #3}
\def\PNAS #1 #2 #3{{\sl Proc. Natl. Acad. Sci. USA} {\bf #1} (#2) #3}
\def\RMP #1 #2 #3{{\sl Rev. Mod. Phys.} {\bf #1} (#2) #3}
\def\PR #1 #2 #3{{\sl Phys. Reports} {\bf #1} (#2) #3}
\def\AoM #1 #2 #3{{\sl Ann. of Math.} {\bf #1} (#2) #3}
\def\UMN #1 #2 #3{{\sl Usp. Mat. Nauk} {\bf #1} (#2) #3}
\def\FAP #1 #2 #3{{\sl Funkt. Anal. Prilozheniya} {\bf #1} (#2) #3}
\def\FAaIA #1 #2 #3{{\sl Functional Analysis and Its Application} {\bf
#1} (#2) #3}
\def\BAMS #1 #2 #3{{\sl Bull. Am. Math. Soc.} {\bf #1} (#2)
#3} \def\TAMS #1 #2 #3{{\sl Trans. Am. Math. Soc.} {\bf #1} (#2) #3}
\def\InvM #1 #2 #3{{\sl Invent. Math.} {\bf #1} (#2) #3}
\def\LMP #1 #2 #3{{\sl Letters in Math. Phys.} {\bf #1} (#2) #3}
\def\IJMPA #1 #2 #3{{\sl Int. J. Mod. Phys.} {\bf A#1} (#2) #3}
\def\AdM #1 #2 #3{{\sl Advances in Math.} {\bf #1} (#2) #3}
\def\RMaP #1 #2 #3{{\sl Reports on Math. Phys.} {\bf #1} (#2) #3}
\def\IJM #1 #2 #3{{\sl Ill. J. Math.} {\bf #1} (#2) #3}
\def\APP #1 #2 #3{{\sl Acta Phys. Polon.} {\bf #1} (#2) #3}
\def\TMP #1 #2 #3{{\sl Theor. Mat. Phys.} {\bf #1} (#2) #3}
\def\JPA #1 #2 #3{{\sl J. Physics} {\bf A#1} (#2) #3}
\def\JSM #1 #2 #3{{\sl J. Soviet Math.} {\bf #1} (#2) #3}
\def\MPLA #1 #2 #3{{\sl Mod. Phys. Lett.} {\bf A#1} (#2) #3}
\def\JETP #1 #2 #3{{\sl Sov. Phys. JETP} {\bf #1} (#2) #3}
\def\JETPL #1 #2 #3{{\sl  Sov. Phys. JETP Lett.} {\bf #1} (#2) #3}
\def\PHSA #1 #2 #3{{\sl Physica} {\bf A#1} (#2) #3}
\def\CQG #1 #2 #3{{\sl Class. Quantum Grav.} {\bf #1} (#2) #3}
\def\SJNP #1 #2 #3{{\sl Sov. J. Nucl. Phys. (Yadern.Fiz.)} {\bf #1} (#2) #3}

\newcommand{\p}[1]{(\ref{#1})}
\begin{document}
\thispagestyle{empty}
\renewcommand{\thefootnote}{\fnsymbol{footnote}}
\begin{flushright}
Preprint DFPD 97/TH/1\\
hep-th/9701037\\
January 1997
\end{flushright}

\bigskip
\begin{center}
{\large\bf Covariant Action for a D=11 Five--Brane
with the Chiral Field}

\vspace{1cm}
Paolo Pasti$^1$\footnote{e--mail: pasti@pd.infn.it},
Dmitri Sorokin$^2$\footnote{e--mail:
kfti@rocket.kharkov.ua, sorokin@pd.infn.it} and
Mario Tonin$^1$ \footnote{e--mail: tonin@pd.infn.it}

\vspace{0.5cm}
$^1${\it Universit\`a Degli Studi Di Padova,
Dipartimento Di Fisica ``Galileo Galilei''\\
ed INFN, Sezione Di Padova,
Via F. Marzolo, 8, 35131 Padova, Italia}

\bigskip
$^2${\it National Science Center\\
Kharkov Institute of Physics and Technology,\\
Kharkov, 310108, Ukraine}

\vspace{1.cm}
{\bf Abstract}
\end{center}

We propose a complete Born--Infeld--like action for a
bosonic 5--brane with a worldvolume chiral field in a background of
gravitational and antisymmetric gauge fields of $D=11$ supergravity.
When the five--brane couples to a
three--rank antisymmetric gauge field, local symmetries
of the five--brane require the addition to the action of an appropriate
Wess--Zumino term. To preserve general coordinate and Lorentz invariance
of the model we introduce a single auxiliary scalar field. The auxiliary
field can be eliminated by gauge fixing a corresponding local symmetry at
the price of the loss of manifest $d=6$ worldvolume covariance. The double
dimensional reduction of the five--brane model results in the Born--Infeld
action with the Wess--Zumino term for a $D=10$ four--D--brane.



\renewcommand{\thefootnote}{\arabic{footnote}}
\newpage

\section{Introduction}
New types of supersymmetric p--branes have attracted recently a great
deal of attention in the course of studying dualities in string theory.
Among them are Dirichlet branes (or D--branes) \cite{le,po} and an
M--theory \cite{M} five--brane \cite{ds}  carrying in its worldvolume a
self--dual (or chiral) two--form gauge field \cite{gt}.  Complete
$\kappa$--invariant actions for super--D--branes were constructed very
recently in \cite{c}, while getting the action for the M--theory
five--brane remains a challenging problem.  Its solution would allow to
shed new light on the structure of D=11 M--theory itself.  Equations of
motion of a  D=11 super--five--brane in a superfield form were obtained in
\cite{hs} by use of a doubly supersymmetric geometrical approach to
describing extended objects \cite{bpstv}, while only partial results has
been obtained about the structure of bosonic part of the
super--five--brane action \cite{5,ah,w,ps}. The action should be of a
higher order in the field strength of the chiral field.

An obstacle to get such an action is caused by the presence of the
second--rank antisymmetric gauge field whose field strength is self--dual
in the free field limit. When one tries to incorporate the self--duality
condition into an action a problem arises with preserving manifest Lorentz
invariance of the model. This problem has a rather long history and
originates from the electric--magnetic duality and the Dirac monopole
problem in Maxwell theory (see \cite{bs} and references there in). Rather
extensive literature has been devoted to the problem of Lorentz covariance of
self--dual field models \cite{fj}--\cite{gir} in connection with their
important role in multidimensional supergravity and string theory. A
non--manifestly Lorentz covariant formulation of chiral boson actions has
been developed in \cite{fj,hen,ss}.  As a generalization of this
approach a d=6 Born--Infeld--like action for a self--interacting chiral
two--form was proposed in \cite{ps} as a base for the construction of the
M--theory five--brane action. The lack of manifest Lorentz invariance will
result in the lack of manifest general coordinate invariance of the
worldvolume of the five--brane, which can substantially complicate the
construction and study of the complete super--five--brane action in such a
formulation.  Thus a covariant formulation which would reproduce the
results of \cite{ps} in a noncovariant gauge is desirable.

The purpose of the present article is to present such a formulation.
We propose a $d=6$ worldvolume covariant action of a
Born--Infeld type for a five--brane with a chiral boson in its
worldvolume.  The action is a generalization of a Lorentz covariant
formulation of duality symmetric and self--dual fields proposed in
\cite{pst2,pst3}.  In contrast to a covariant formulation of chiral bosons
with infinitely many auxiliary fields \cite{mwy,wot,dh,bk,ber1,ber2} in
our approach $d=6$ covariance is achieved by introducing a single
auxiliary scalar field entering the action in a nonpolynomial way
\cite{pst2,pst3}
\footnote{See ref.  \cite{pst3} on a relation between the
two approaches and ref. \cite{ben} for the application of the approach
with infinitely many fields to studying the duality of Born--Infeld
actions.}.  The auxiliary field ensures not only worldvolume covariance but
also that all the constraints of the model are of the first class \cite{pst3}
which is important for performing covariant quantization.  Upon eliminating
the auxiliary scalar field by a noncovariant gauge fixing of a local symmetry
the action reduces (in the case of the flat worldvolume metric) to the
non--manifestly Lorentz invariant action of \cite{ps} describing a
self--interacting second--rank chiral field.  Since the bosonic p-branes can
live in target space of any dimension higher than $D=p+1$ we do not specify
the dimension of the target space of the five--brane at hand, but imply that
$D=11$ having in mined possible supersymmetric generalization of the model.

We couple the five--brane to antisymmetric gauge fields of $D=11$
supergravity. A remarkable feature of the model is that when the
five--brane couples to a three--rank antisymmetric gauge field, local
symmetries of the five--brane require the addition to the action of an
appropriate Wess--Zumino term.  Upon the double dimensional reduction of
the five--brane worldvolume from $d=6$ to $d=5$ and target space from
$D=11$ to $D=10$ the five--brane action reduces to the Born--Infeld action
with a Wess--Zumino term for a Dirichlet four--brane.

\section{The action}

Consider a five--brane described by the following action invariant under
$d=6$ general coordinate transformations:
\begin{equation}\label{action}
S=\int d^6x\left[\sqrt{-g}{1\over
{4\partial_r a\partial^r a}}\partial_ma(x)F^{*mnl}F_{nlp}\partial^pa(x)+
\sqrt{-\det(g_{mn}+i\tilde F_{mn})}\right],
\end{equation}
where $x^m$ (m=0,1...,5) are
local coordinates of the worldvolume,\\
$g_{mn}(x)=\partial_mX^M(x)g_{MN}\partial_nX^N(x)$ is a worldvolume metric
induced by embedding into curved target space with the metric $g_{MN}(X)$
parametrized by coordinates $X^M$ (M,N=0,...,D-1);
$F_{mnl}=2(\partial_{l}A_{mn}+
\partial_{m}A_{nl}+\partial_{n}A_{lm})$ is the field
strength of an antisymmetric worldvolume gauge field $A_{mn}(x)$;
$g=\det{g_{mn}}$; $F^{*lmn}$ is the dual field strength:
$$ F^{*lmn}={1\over {6\sqrt{-g}}} \varepsilon^{lmnpqr}F_{pqr} $$
and
$$\tilde
F_{mn}\equiv{1\over{\sqrt{\partial a)^2}}}F^{*}_{mnl}\partial^la(x).$$

The scalar field $a(x)$ ensures manifest $d=6$ covariance of the model and
is completely auxiliary, as we shall see below.

Note that in spite of the presence of the imaginary unit inside the
determinant in \p{action} the letter is real. This can be seen by
rewriting the determinant as a polynomial in powers of $\tilde F$ (as in
\cite{ps})
\begin{equation}\label{det}
\det(g_{mn}+i\tilde F_{mn})=g(1+{1\over 2}tr\tilde F^2+
{1\over 8}(tr\tilde F^2)^2-{1\over 4}tr\tilde F^4).
\end{equation}
Though the argument of the determinent is $6\times 6$ matrix the
polynomial in the r.h.s.  of \p{det} stops at the 4-th power of $\tilde F$
since (by construction) $\tilde F_{mn}$ is degenerate and has rank 4.

If in \p{action} we take the flat metric and
insert the term $-{1\over 2}\tilde F_{mn}\tilde F^{mn}$
instead of the square root the resulting
action
$$
S=\int d^6x{1\over
{4(\partial a)^2}}\partial_ma(x)F^{*mnl}(F_{nlp}-F^*_{nlp})\partial^pa(x)
$$
\begin{equation}\label{fine}
\equiv
\int d^6x [{1\over 24}F_{lmn}F^{lmn}
-{1\over{8(\partial
a^2)}}\partial^ma{(F-F^*)}_{mnl}{(F-F^*)}^{nlr}\partial_ra].
\end{equation}
will describe a free self--dual field $A_{mn}$ with
\begin{equation}\label{sd}
F_{mnl}-F^*_{mnl} =0
\end{equation}
(see \cite{pst3} for the details).

The action \p{action} is invariant under worldvolume diffeomorphisms and the
following local transformations:
\begin{equation}\label{ordi}
\delta
A_{mn}=\partial_{[m}\phi_{n]}(x),
\end{equation}
(which is the ordinary gauge
symmetry of the massless antisymmetric fields),
\begin{equation}\label{varm}
\delta A_{mn}={1\over 2}\partial_{[m}a(x)\varphi_{n]}(x), \qquad
\delta a(x)=0 \end{equation}
and
$$ \delta a(x)=\varphi(x), $$
\begin{equation}\label{phi}
\delta A_{mn}={{\varphi(x)}\over {2(\partial
a)^2}}(F_{mnp}\partial^pa-{\cal V}_{mn}),
\end{equation} where
$$ {\cal
V}^{mn}\equiv-2\sqrt{-{{(\partial a)^2}\over{g}}}
{{\delta{\sqrt{-\det(g_{pq}+i\tilde F_{pq})}}}\over{\delta\tilde F_{mn}}}.
$$
(The definition of ${\cal V}^{mn}$ is chosen in such a way that in the
free limit \p{fine} it coincides with
$\tilde F^{mn}$.)

The transformations \p{varm} will allow us to algebraically eliminate part
of components of the chiral field $A_{mn}$. The invariance of the action
\p{action} under \p{ordi} and \p{varm} is obvious, and the variation of
the action under the transformations \p{phi} is
\begin{equation}
\label{0} \delta S=\int d^6x{1\over 2}
\left[\sqrt{-{{g}\over{(\partial
a)^2}}}(F_{mnp}\partial^p a - {\cal V}_{mn})\delta(\tilde F^{mn})
+{\sqrt{-g}\over{(\partial a)^2}}F^{*mnp}F^{~~q}_{mn}
\partial_{[p}a\partial_{q]}\varphi\right]=0,
\end{equation}
where the variation of $\tilde F^{mn}$ includes the
variation of $A_{mn}$ and $a(x)$ with the parameter $\varphi(x)$ \p{phi}.
The proof that \p{0} is zero (up to a total derivative) can be performed
along the same lines as in \cite{ps}, where a non--covariant action for a
self--interatcting chiral field was demonstrated to possess a modified
non--manifest Lorentz invariance.

The equations of motion of $A_{mn}$, which follow from \p{action} are
\begin{equation}\label{A}
\varepsilon^{lmnpqr}\partial_n{{\partial_pa}\over{(\partial a)^2}}
(F_{qrs}\partial^sa-{\cal V}_{qr})=0.
\end{equation}
An appropriate gauge fixing of transformations \p{varm} allows one
\cite{pst2,pst3} to reduce the general solution of \p{A} to the form
\begin{equation}\label{gs}
F_{qrs}\partial^sa-{\cal V}_{qr}=0
\end{equation}
which is a generalization of the
self--duality condition \p{sd} to the case of the self--interacting field
$A_{mn}$.

As in the free field case \cite{pst2,pst3}, the equation of motion
of $a(x)$ turns out to be a consequence of \p{A} and, hence, is not a new
field equation. This permits, without losing information on the
dynamics of the model, to eliminate $a(x)$ directly from the action by
gauge fixing transformations \p{phi} \footnote{Note that the gauge
$\partial_ma\partial^ma=0$ (such as, for instance, $a=const$) is
inadmissible because of the presence of $(\partial a)^2$ in the
denominator of Eq. \p{action}. This would lead to a singularity of the
action.}.  However, the price for such a gauge fixing is the loss of
worldvolume general coordinate invariance, or Lorentz invariance of the
model in the flat limit.  For instance, in the case of the flat $d=6$
metric by putting
\begin{equation}\label{gf}
\partial_ma(x)=\delta^5_m
\end{equation}
to be the unit vector along the fifth spatial direction of
the worldvolume we reproduce the action and equations of motion of a
self--interacting chiral field $A_{mn}$ constructed in \cite{ps}.  Note
that in the gauge \p{gf} we can also completely eliminate the components
$A_{m5}$ of the gauge field by use of the algebraic local transformations
\p{varm}:
\begin{equation}\label{a5} A_{m5}=0.
\end{equation}
Then the
modified Lorentz transformations of remaining components $A_{\alpha\beta}$
$(\alpha, \beta =0,1,...,4)$ of Ref. \cite{ps} arise in our approach as a
combination (which
preserves the gauge fixing condition \p{gf}) of the standard Lorentz
transformations with parameters $\Lambda_{mn}=-\Lambda_{nm}$ and the
transformation \p{phi} :
\begin{equation}\label{combi}
\delta(\partial_ma(x))=\Lambda_m^{~n}\partial_na+x^p\Lambda_p^{~n}\partial_n
(\partial_ma)+\partial_m\varphi(x)=\Lambda_m^{~5}+\partial_m\varphi(x)=0.
\end{equation}
From \p{combi} it follows that to preserve \p{gf} the parameter $\phi(x)$
of \p{phi} must be of the form:
\begin{equation}\label{par}
\phi(x)=-x^m\Lambda_m^{~5}=-x^\alpha\Lambda_\alpha^{~5}.
\end{equation}

Substituting \p{gf} and \p{par} into \p{phi} and combining it with the
Lorentz transformation mixing $\alpha=(0,1,...,4)$ directions with the 5
direction we get for $A_{\alpha\beta}$ (in the gauge $A_{m5}=0$):
$$
\delta
A_{\alpha\beta} = (x^\gamma \Lambda_\gamma^{~5})\partial_5A_{\alpha\beta}-
x^5(\Lambda_\gamma^{~5}\partial_\gamma)A_{\alpha\beta}-
x^\alpha\Lambda_\alpha^{~5}(\partial_5A_{\alpha\beta}-{1\over 2}
{\cal V}_{\alpha\beta})
$$
\begin{equation}\label{modi}
=(x^\gamma \Lambda_\gamma^{~5}){1\over 2}{\cal
V}_{\alpha\beta}-
x^5(\Lambda_\gamma^{~5}\partial^\gamma)A_{\alpha\beta},
\end{equation}
which are exactly the modified Lorentz transformations of Ref. \cite{ps}.
They coincide with ordinary Lorentz transformations on the mass shell
\p{gs}. Thus, though upon imposing the gauge fixing condition \p{gf} the
generalized self--duality condition loses the manifestly Lorentz covariant
form, it is nevertheless invariant under the Lorentz transformations
\p{modi}.

\section{Coupling to D=11 antisymmetric gauge fields}
Consider now the propagation of the five--brane in a background of
antisymmetric gauge fields of D=11 supergravity. These are a three--rank
field $C^{(3)}_{LMN}(X)$ and its dual six--rank field
$C^{(6)}_{LMNPQR}(X)$ \cite{cl}.
The coupling requires the replacement of
the field strength $F_{lmn}(x)$ with
\begin{equation}\label{h}
H_{lmn}=F_{lmn}-C^{(3)}_{lmn}
\end{equation}
and adding to the action \p{action}
a Wess--Zumino term \cite{ah}. The resulting action becomes
$$
S=\int dx^6\left[
\sqrt{-det(g_{mn}+i\tilde H_{mn})}+\sqrt{-g}{1\over {4\partial_r
a\partial^r a}}\partial_ma(x)H^{*mnl}H_{mnp}\partial^pa(x)\right]
$$
\begin{equation}\label{ac}
+\int\left[C^{(6)}+{1\over 2}F\wedge C^{(3)}\right],
\end{equation}
where  in \p{h} and \p{ac} the forms $C^{(6)}$ and $C^{(3)}$ are pullbacks
into $d=6$ worldvolume of the corresponding $D=11$ forms. The last two
terms of \p{ac} form the Wess--Zumino term.
The coefficient in front of the last term in \p{ac}
is singled out by the requirement that the action \p{ac} remains invariant
under the transformations \p{varm} and \p{phi}, where in the latter $F$ is
replaced with $H$ \p{h}. The transformations of the last term in \p{ac}
compensate part of the transformations of the second $(H^*H)$ term while
the remaining ones are canceled by the corresponding transformations of
the Born--Infeld--like part of the action (like in the absence of the
background fields). Thus the Wess--Zumino term is required to preserve
local symmetries of the action when the five--brane couples to the
antisymmetric fields.

Since $C^{(3)}$ and $C^{(6)}$ are $D=11$ gauge fields transformed as
\begin{equation}\label{chi}
\delta C^{(3)}(X)=2d\chi(X), \qquad  \delta C^{(6)}(X)=-
d\chi(X)\wedge C^{(3)}(x),
\end{equation}
for the action \p{ac} to be
invariant under \p{chi} the chiral field $A_{mn}$ must transform as:
$$
\delta A(x)=\chi\left(X(x)\right).
$$

This completes the construction of the action for the bosonic 5--brane
propagating in the gravitational and gauge field background of
11--dimensional supergravity.

\section{Reduction to a four--D--brane in D=10}.

Let us perform a double dimensional reduction
of the action \p{ac} to a four--brane propagating in a D=10
background. This procedure consists in wrapping $x^5$ dimension of the
five--brane around $X^{10}$ and requiring that all the fields of the
reduced model do not depend on $x^5$ and $X^{10}$. Below we shall consider
a simplified variant of the dimensional reduction by putting to zero
dilaton and vector field which arise from components of the dimensionally
reduced metrics and postpone a complete and more systematic analysis to
more detailed future paper.

To carry out the dimensional reduction we first put the gauge conditions
\p{gf} and \p{a5}, split the $d=6$ indices into $d=5$ and the 5-th ones
and drop the index 5. Thus instead of the original three-- and six--rank
fields we have:
\begin{equation}\label{split1}
F_{lmn}~~~\Rightarrow~~~(F_{\alpha\beta\gamma},~F_{\alpha\beta}),
\end{equation}
\begin{equation}\label{split2}
C^{(3)}_{lmn}~~~\Rightarrow~~~(C^{(3)}_{\alpha\beta\gamma},
~C^{(2)}_{\alpha\beta}), \qquad C^{(6)}~~~\Rightarrow~~~{6}C^{(5)}.
\end{equation}
Note that because of the independence of $x^5$
and by virtue of \p{a5} the field strength $F_{\alpha\beta}$ in \p{split1}
is zero, and only the field $A_{\alpha\beta}$ live in the $d=5$
worldvolume.

In terms of \p{split1} and \p{split2} the reduced action takes the form
$$
S=\int dx^5\left[
\sqrt{-\det(g_{\alpha\beta}+iH^*_{\alpha\beta})}+{{\sqrt{-g}}\over
4}H^{*\alpha\beta}H_{\alpha\beta}\right]
$$
\begin{equation}\label{red}
+\int dx^5\varepsilon^{\alpha\beta\gamma\delta\sigma}
\left[C^{(5)}_{\alpha\beta\gamma\delta\sigma}+{1\over 24}
F_{\alpha\beta\gamma}C^{(2)}_{\delta\sigma}\right],
\end{equation}
where $H^{*\alpha\beta} =-{1\over {6\sqrt{-g}}}
\varepsilon^{\alpha\beta\gamma\delta\sigma
5}(F-C^{(3)})_{\gamma\delta\sigma}$.

We see that the second term of \p{ac}  became of the Wess--Zumino
type in \p{red} and contributes to the Wess--Zumino term, the resulting
action being of the form:
$$
S=\int dx^5
\sqrt{-\det(g_{\alpha\beta}+iH^*_{\alpha\beta})}+
$$
\begin{equation}\label{res}
+\int dx^5\varepsilon^{\alpha\beta\gamma\delta\sigma}
\left[C^{(5)}_{\alpha\beta\gamma\delta\sigma}+
{1\over 12}F_{\alpha\beta\gamma}C^{(2)}_{\delta\sigma}-
{1\over 24}C^{(3)}_{\alpha\beta\gamma}C^{(2)}_{\delta\sigma}\right].
\end{equation}
The action \p{res} is a dual form of the action for a four--D--brane
\cite{ts}. It can be rewritten in the conventional Born--Infeld form
involving the field strength $\hat F_{\alpha\beta}$ of a vector field
\cite{bt}
by
performing a dualization procedure inverse to that described in
\cite{ts}, so we shall not discuss this point in detail but just note that
formally it consists in replacing $F_{\alpha\beta\gamma}$,
$C^{(2)}_{\delta\sigma}$ and $C^{(3)}_{\alpha\beta\gamma}$ with their duals:
$$iF^*_{\alpha\beta}~~\rightarrow~~\hat F_{\alpha\beta} \qquad
iC^{(3)*}_{\alpha\beta}~~\rightarrow~~
\hat C^{(2)}_{\alpha\beta}\qquad
C^{(2)}_{\alpha\beta}~~\rightarrow~~12i\hat C^{(3)*}_{\alpha\beta},
$$
and adjusting coefficients in an appropriate way required by the strict
dualization procedure \cite{ts}.
As a result we get
$$ S=\int dx^5
\sqrt{-\det(g_{\alpha\beta}+(\hat F-\hat C^{(2)})_{\alpha\beta})}+
$$
\begin{equation}\label{resa}
+\int dx^5\varepsilon^{\alpha\beta\gamma\delta\sigma}
\left[C^{(5)}_{\alpha\beta\gamma\delta\sigma}+
(\hat F-\hat C^{(2)})_{\alpha\beta}\hat
C^{(3)}_{\gamma\delta\sigma}-{1\over 2}\hat C^{(2)}_{\alpha\beta} \hat
C^{(3)}_{\gamma\delta\sigma}\right].
\end{equation}
The last term of \p{resa} can be included into a redefined
$\hat C^{(5)}$ and
then the Wess--Zumino term in \p{resa} take a canonical formal form
$exp(\hat F-\hat C^{(2)})(\hat C^{(5)}+\hat C^{(2)})$ \cite{bt}.

\section{Conclusion}
We have constructed the complete Born--Infeld--like action for the
bosonic 5--brane carrying the worldvolume chiral field in a background of
gravitational and antisymmetric gauge fields of $D=11$ supergravity.  To
preserve general coordinate and Lorentz invariance of the model we
introduced the single auxiliary scalar field. The auxiliary field can be
eliminated by gauge fixing the corresponding local symmetry at the price
of the loss of manifest space--time covariance. The double dimensional
reduction of the five--brane model results in the Born--Infeld
action for the $D=10$ four--D--brane.

The form of the covariant action \p{ac}
and its relation to the D--brane action
allows one to hope that it can be
generalized to describe embedding of five--brane worldvolume into a
superfield background of $D=11$ supergravity. In other words
it should be possible to construct a $d=6$ covariant $\kappa$--symmetric
action for a super--five--brane of M--theory analogous to that found for
the super--D--branes \cite{c}, and which may have the form of \p{ac}
with all background fields replaced with corresponding superfields.  It
would be also of interest to compare the five--brane equations of motion
obtained from such an action with covariant superfield equations for the
M--theory five--brane proposed in \cite{hs}, in particular, to find out if
there is an analog of the auxiliary scalar field being crucial for the
covariance of our model.  Other interesting problems are to relate by a
dimensional reduction the M--theory five--brane to a type IIA five--brane
and a heterotic five--brane in ten dimensions. Work in these directions is
in progress.

\bigskip
\noindent
{\bf Acknowledgements}. The authors are grateful to I. Bandos, C.
Preitschopf and K. Lechner for illuminating discussion.  This work was
supported by the European Commission TMR programme ERBFMRX--CT96--045 to
which P.P. and M.T. are associated.  D.S. acknowledges partial support
from the INTAS Grants N 93--127 and N 93--493.

\medskip
\noindent
{\bf Note added}.
After this paper was sent to hep-th and submitted for publication the
authors learned about an article by J. H. Schwarz \cite{jhs} where the
chiral tensor field was coupled to $d=6$ gravity in the noncovariant
formulation.  When the antisymmetric field background is switch off and
the auxiliary field is gauge fixed by Eq. \p{gf} our model reduces to that
of \cite{jhs}.

\end{document}